\documentclass[preprint]{revtex4}

\begin{document}

\title{The cosmic coincidence in Brans-Dicke cosmologies}

\author{Saulo Carneiro\footnote{This essay received an ``honorable mention" in the 2005
Essay Competition of the Gravity Research Foundation.}}

\affiliation{Instituto de F\'{\i}sica, Universidade Federal da
Bahia, 40210-340, Salvador, BA, Brazil}

\begin{abstract}
Among the suggested solutions to the cosmological constant
problem, we find the idea of a dynamic vacuum, with an energy
density decaying with the universe expansion. We investigate the
possibility of a variation in the gravitational constant as well,
induced, at the cosmological scale, by the vacuum decay. We
consider an effective Brans-Dicke theory in the spatially flat
FLRW spacetime, finding late time solutions characterized by a
constant ratio between the matter and vacuum energy densities. By
using the observed limits for the universe age, we fix the only
free parameter of our solutions, obtaining a relative matter
density in the range $0.25<\Omega_m<0.4$. In particular, for
$Ht=1$ we obtain $\Omega_m=1/3$. This constitutes a possible
explanation for another problem related to the cosmological term,
the cosmic coincidence problem.
\end{abstract}

\maketitle

We are celebrating the 90th birthday of General Relativity, a
theory that has survived to several precise tests, including
verifications of the Equivalence Principle in its weak and strong
versions. Such tests generally refer to weak or intermediate
gravitational fields, but we have enough reasons to believe that
General Relativity is the classical theory of gravitation also in
the strong field limit.

The theory has then the same theoretical and observational status
of Maxwell's electromagnetism. Nevertheless, quantum
electrodynamics predicts corrections to its classical version in
the presence of very strong fields, corrections related to vacuum
effects, as the running of masses and couplings with the energy
scale, and the break of the linearity (the principle of
superposition) characteristic of classical electrodynamics.

Despite the absence of a quantum version of General Relativity,
one can argue about similar vacuum effects in the strong field
limit. In this case, a running of masses would be equivalent (at
least in what concerns gravitation) to a running of the
gravitational constant $G$ (breaking the strong version of the
equivalence principle).

Consider, for example, the problem of vacuum energy in cosmology.
When we derive the vacuum energy density by means of quantum field
theories in flat spacetime, we obtain a divergent result, which
must be exactly canceled by a bare cosmological constant in
Einstein's equations, because in the flat spacetime the left-hand
side of these equations is identically zero. Now, if we calculate
the vacuum density in the expanding background, we also obtain a
divergent result, but a renormalized value may be derived by
subtracting the Minkovskian contribution. The resulting vacuum
density will depend on the curvature, and so will vary with time
as the universe expands. This mechanism may constitute a possible
explanation for the cosmological constant problem
\cite{Starobinsky,Peebles}, leading from a huge value at Planck
times to the small value observed nowadays \cite{Bertolami,OT}.

In the absence of a definite quantum field theory in curved
spacetimes, the effects of a decaying vacuum density and of an
induced time variation of $G$ may be investigated in the realm of
effective geometric theories that respect the energy conservation
and the weak equivalence principle. An example is the
scalar-tensor Brans-Dicke theory \cite{BD}, which will be
considered here (for another approach, see \cite{MGX,Friedmann}).
As ``effective" we mean, among other things, that the parameter
$\omega$ characteristic of the theory depends on the scale, being
very high in the weak field case (leading to unperturbed Einstein
equations), but of the order of unity in the cosmological limit.
As we are concerned only to this limit, we will use the simplest
theory, with constant $\omega$.

Note that we are considering a time variation of $G$ at the
cosmological scale, and therefore such a variation cannot be ruled
out by local observations. In the case of a stationary spacetime,
$G$ or any other dynamical quantity are independent on time. (Even
in the context of Brans-Dicke theories, the scalar field $\phi$
has different dynamics at the local and cosmological scales, in
the same way as the metric tensor field.)

Consider a spatially flat FLRW spacetime filled with a cosmic
fluid composed by dust with energy density $\rho_m$, and by a
vacuum term with energy density $\rho_{\Lambda}$ and pressure
$p_{\Lambda} = - \rho_{\Lambda}$. This equation of state for the
vacuum is a natural choice, since the vacuum has the symmetry of
the background. Then, the total energy density is given by $\rho =
\rho_m + \rho_{\Lambda}$, while the total pressure is $p = -
\rho_{\Lambda}$. Under these conditions, the Brans-Dicke field
equations can be written as
\begin{equation} \label{1}
H^2 =
\frac{8\pi}{3}\frac{\rho}{\phi}-\frac{\dot{\phi}}{\phi}H+\frac{\omega}{6}
\frac{\dot{\phi}^2}{\phi^2},\end{equation} \begin{equation}
\label{2} \dot{\rho}=-3H(\rho+p)=-3H\rho_m,\end{equation}
\begin{equation} \label{3}
\frac{d}{dt}(\dot{\phi}a^3)=\frac{8\pi}{3+2\omega}(4\rho-3\rho_m)a^3,
\end{equation}
where $H=\dot{a}/a$ is the Hubble parameter.

If $\rho_{\Lambda}$ is constant, this system is solvable. But, in
the case of a decaying vacuum density, we have three equations to
determine the functions $a$, $\rho$, $\rho_m$ and $\phi$.
Therefore, it is necessary to add some physical constraint, which
may be the evolution law for $\rho_{\Lambda}$, or for $G$, which
could be obtained if we had a definite description of vacuum in
expanding backgrounds. Here we will infer an evolution law for $G$
from an observed relation between $G$ and $H$ (usually called
Eddington-Dirac or Weinberg relation), given by $G \approx H/m^3$,
where $m$ is the energy scale of the QCD phase transition, the
latest cosmological vacuum transition. We have presented elsewhere
a possible justification for that relation, based on the
holographic conjecture \cite{Friedmann,GRF,GS}. Here we will
simply postulate that, if it is valid nowadays, it may also be
valid at any time of the present phase of universe evolution. As,
in Brans-Dicke theory, $G\sim\phi^{-1}$, we then write
\begin{equation} \label{EW}
\phi = \frac{G_0}{G}=\frac{8\pi G_0\lambda}{H},
\end{equation}
where $G_0$ and $\lambda$ are positive constants of the order of
unity and $m^3$, respectively.

With this constraint one can obtain the general solution of
(\ref{1})-(\ref{3}). For our purposes it is enough to find a class
of particular solutions describing the late time observed
universe. Observations suggest that $\rho_m$ is approximately one
third of the critical density $\rho_c=3H^2/(8\pi G)$ (note that,
in the case of Brans-Dicke theories, the flatness of space does
not imply $\rho=\rho_c$). On the other hand, the vacuum energy
density contributes, at most, with a similar figure, otherwise its
effects would be more evident. Therefore, the total energy density
is close to $\rho_c$. As before, we will assume that this is not a
mere coincidence, but it is a functional relation, valid for any
time of the latest phase of the expansion. With the help of
(\ref{EW}) we can then write
\begin{equation} \label{rho}
\rho = \frac{3\gamma H^2}{8\pi G} = 3\gamma\lambda H,
\end{equation}
with $\gamma\sim1$.

Before going on, let us stress that relations (\ref{EW}) and
(\ref{rho}) are supposed valid during the present epoch, that is,
in the limit of late times. The variation law for $G$ (or for the
vacuum density) in each era of universe history will depend on the
underlying vacuum physics, and is not necessarily the same
throughout the whole expansion. Therefore, the model presented
here cannot say, in its present form, anything concerning earlier
time phenomena like nucleosynthesis, the cosmic background
radiation or structure formation.

Substituting the ansatz (\ref{EW})-(\ref{rho}) in equation
(\ref{1}), we obtain an evolution equation for $H$, given by
\begin{equation} \label{H}
AH^4-\dot{H}H^2-B\dot{H}^2=0,
\end{equation}
where
\begin{eqnarray}\label{A}
A &\equiv& 1 - \frac{\gamma}{G_0}, \\ \label{B} B &\equiv&
\frac{\omega}{6}.
\end{eqnarray}

For $A=0$ (that is, $\gamma/G_0=1$), the de Sitter universe is a
possible solution. Apart this case, the general solution of
(\ref{H}) is given by
\begin{equation} \label{age}
Ht = n,
\end{equation}
with
\begin{equation} \label{n}
n = \frac{\sqrt{1+4AB}-1}{2A},
\end{equation}
and where the integration constant was chosen so that the
divergence of $H$ occurs at $t=0$.

Integrating once more we obtain, for the scale factor,
\begin{equation}\label{a}
a(t) = a_0 t^n.
\end{equation}
The deceleration parameter $q \equiv - a\ddot{a}/\dot{a}^2$ is
then given by
\begin{equation} \label{q}
q = \frac{1}{n}-1.
\end{equation}

Let us now verify whether this solution satisfies the remaining
field equations. Substituting (\ref{age}) and (\ref{rho}) into
(\ref{2}) leads to
\begin{equation}\label{rho_m}
\rho_m = \frac{\lambda\gamma}{n}H.
\end{equation}
From this and (\ref{rho}), one obtains the relative energy density
of matter,
\begin{equation}\label{Omega_m}
\Omega_m \equiv \frac{\rho_m}{\rho} =  \frac{1}{3n}.
\end{equation}

This is an interesting result, meaning that, in the present phase
of universe evolution, the relative matter density is constant.
This may solve a second problem related to the cosmological
constant, namely the approximate coincidence between the matter
and vacuum energy densities. Here, such a coincidence follows
naturally, without necessity of any fine tuning.

Note also that our solution depends on just one parameter, the
constant $n$ in (\ref{age}) (usually called the age parameter).
So, fixing $n$ by using, say, observational limits for the
universe age, we can verify whether the solution agrees with other
cosmological parameters, as the relative matter density or the
deceleration parameter.

Current astrophysical limits on the universe age lead to an age
parameter in the interval $0.8< Ht < 1.3$ \cite{age}. Then, from
(\ref{age}) and (\ref{Omega_m}) we obtain $0.25<\Omega_m<0.4$. The
most probable value $Ht\approx1$ leads to $\Omega_m\approx1/3$. On
the other hand, from (\ref{q}) we have $-1/4<q<1/4$, the most
probable value being $q\approx0$. The observations still are not
sufficiently precise to give the correct value of $q$.
Nevertheless, some authors have recently claimed that, if the
scale factor follows a potential law like (\ref{a}), the best
fitting of redshift-distance relations for supernova Ia and for
compact radius sources is obtained for $n\approx1$, that is, for
$q\approx0$ \cite{Dev,Alcaniz}.

From (\ref{rho_m}), (\ref{rho}) and (\ref{EW}), it follows, for
the vacuum term,
\begin{equation} \label{Schutz}
\rho_{\Lambda} =\rho - \rho_m = \left( 3-\frac{1}{n} \right)
\gamma \lambda H \sim m^3 H,\end{equation} \begin{equation}
\label{Lambda} \Lambda = 8\pi G \rho_{\Lambda} = \left(
3-\frac{1}{n} \right) \gamma H^2.
\end{equation}
We obtain a decaying vacuum density, in the spirit of the
introduction, with no necessity of postulating a variation law in
our original ansatz. Surprisingly enough, the above equations
agree with results derived on the basis of quantum field
estimations in expanding backgrounds \cite{Ralf,Shapiro}. The
decaying law (\ref{Lambda}) has also been heuristically suggested
by several authors in both, varying or constant $G$, cases (see,
for example, \cite{Ademir,Ademir2,Aldrovandi}).

The reader may note that the matter density (\ref{rho_m}) does not
scale as $a^{-3}$, as it should if the energy of matter were
conserved. This is consequence of the conservation of the total
energy, expressed by equation (\ref{2}). The vacuum decay is only
possible if associated to a process of matter production (a
general feature of vacuum states in non-stationary spacetimes).
With the help of (\ref{rho_m}), (\ref{a}) and (\ref{age}), it is
possible to verify that the rate of production (the relative
variation rate of $\rho_m a^3$) is given by $\Gamma = (3-1/n)H$.
In what concerns the variation of $G$ (at the cosmological scale),
from equation (\ref{EW}) we have $\dot{G}/G=-(1+q)H$.

Finally, we have to verify equation (\ref{3}). By using
(\ref{EW}), (\ref{rho}), (\ref{age}), (\ref{a}), and
(\ref{rho_m}), it reduces to
\begin{equation} \label{3'}
4n-1 = \frac{3 + 12B}{1-A},
\end{equation}
where $A$ and $B$ were defined in (\ref{A}) and (\ref{B}). The
system of equations (\ref{3'}) and (\ref{n}) allows one to express
$A$ and $B$ (that is, $\omega$ and $\gamma/G_0$) as functions of
the age parameter $n$. For $n\approx1$, one obtains $\omega
\approx 6/5$, and $\gamma/G_0 \approx 9/5$. Therefore, both
$\omega$ and $G_0$ are positive and (since $\gamma \sim 1$) of the
order of unity.

The author is thankful to J. Alcaniz, O. Bertolami, O. Aguiar, G.
Matsas, L.C.B. Crispino, R. Abramo, V. Faraoni, and J.
Garcia-Bellido for useful discussions.

\end{document}